\documentclass{vgtc}                          

\ifpdf
  \pdfoutput=1\relax                   
  \usepackage{graphicx}                
  \DeclareGraphicsExtensions{.pdf,.png,.jpg,.jpeg} 
\else
  \usepackage{graphicx}                
  \DeclareGraphicsExtensions{.eps}     
\fi%

\graphicspath{{figures/}{pictures/}{images/}{./}} 

\usepackage{microtype}                 
\PassOptionsToPackage{warn}{textcomp}  
\usepackage{textcomp}                  
\usepackage{mathptmx}                  
\usepackage{times}                     
\usepackage{cite}                      
\usepackage{tabu}                      
\usepackage{booktabs}                  
\usepackage{xcolor}
\usepackage{svg}
\definecolor{kacolor}{rgb}{.19, .55, .91}
\definecolor{mscolor}{rgb}{.91, .55, .19}

\usepackage[hyphens]{url}
\usepackage{breakurl}
\usepackage{csquotes}
\usepackage{tikz}
\usepackage[final]{changes}
\usepackage{bbding}
\usepackage{dingbat}
\definecolor{solutioncolor}{rgb}{0.01, 0.75, 0.24}
\newcommand{\solution}{\textcolor{black}{\small{\FiveStarOpen}}\hspace{0.3mm}~} 
\definecolor{personalcolor}{rgb}{0.08, 0.38, 0.74}
\definecolor{callcolor}{rgb}{0.7, 0.11, 0.11}
\newcommand{\call}{\textcolor{black} {\small{\PencilRightDown}}\hspace{0.3mm}~} 

\newcommand{\personal}{\textcolor{black}{\small\eye}\hspace{0.2mm}~} 

\onlineid{0}

\vgtccategory{Research}

\vgtcinsertpkg

\title{Toward Inclusion and Accessibility in Visualization Research: Speculations on Challenges, Solution Strategies, and Calls for Action (Position Paper) }
\hypersetup{
           breaklinks=true}  

\author{Katrin Angerbauer\thanks{e-mail: katrin.angerbauer@visus.uni-stuttgart.de} %
\and Michael Sedlmair \thanks{e-mail: michael.sedlmair@visus.uni-stuttgart.de} }
\affiliation{\scriptsize Visualization Research Center (VISUS), University of Stuttgart, Germany}

\abstract{Inclusion and accessibility in visualization research have gained increasing attention in recent years. However, many challenges still remain to be solved on the road toward a more inclusive, shared-experience-driven visualization design and evaluation process. In this position paper, we discuss challenges and speculate about potential solutions, based on related work, our own research, as well as personal experiences. The goal of this paper is to start discussions on the role of accessibility and inclusion in visualization design and evaluation.

} 

\CCScatlist{
  \CCScatTwelve{Human-centered computing}{Visu\-al\-iza\-tion}{Visualization design and evaluation methods}{};
  \CCScatTwelve{Human-centered computing}{Accessibility}{Accessibility design and evaluation methods}{}
}

\begin{document}


 
\maketitle

\section{Introduction} 
The United Nations stresses the need for universal access to information in its convention on the rights of people with disabilities~\cite{UNConvention}.
As visualizations aim to convey information to people, they should also be universally accessible. 
The visualization community is becoming increasingly aware of the need for inclusive and accessible visualizations~\cite{Bongshin2020}. The number of publications with respect to visualization accessibility, though still small, is growing~\cite{Lundgard2019,wu2021,kim2021,Elavsky2022,Joyner2022}.
However, designing and evaluating accessible visualizations remains challenging~\cite{ming2021,Joyner2022}.
There are many pitfalls within the accessibility research process such as parachute research~\cite{Lundgard2019,costanza2020design}, prototypes that do not really benefit users' needs\cite{jackson2022}, and underlying biases~\cite{leal2021activism,mckercher2020beyond}. 

For us visualization researchers, an obvious example would be the focus on visual disabilities, which many have already tried to counteract~\cite{Lundgard2019,kim2021}. But will our field at some point arrive at providing data representations which are truly adaptive to someone's abilities and needs?
There is also a lot of specialized hardware that we might leverage. Virtual and augmented reality could, for instance, provide new forms of accessible data access. However, we also have the responsibility to create augmentation possibilities and virtual worlds in such a way that they are accessible to all~\cite{mott21}.

Since almost two decades, BELIV has been a place to propose and discuss novel evaluation, and more recently, also design methods and processes. Following this year's motto ``Designing and Evaluating Visualizations for an Ethical, Inclusive, and Responsible Future'', in this position paper, 
we provide our thoughts on ability-inclusive, shared-experience-driven visualization evaluation and design processes that define accessibility as a central design choice. Our ideas are based on three pillars: (1) existing literature on accessibility in visualization and other research areas, (2) our own first experiences conducting accessibility research in visualization~\cite{angerbauer2022}, as well as (3) personal experiences and reflections of the first author who has a disability herself.

Based on these three pillars, we formulate challenges on the road toward our envisioned process and speculate about~\solution\textit{potential solutions} for them. We further enrich those with specific ~\call\textit{calls for action}, and~\personal\textit{the personal experience} of the first author. The solutions are by no means exhaustive and usually refer to something that has already been done to attempt to mitigate the problem. Calls for action are starting points for future research efforts. These categories are not clear-cut, as sometimes the mitigation of a problem might also be to do more research.
Further, we use the term \enquote{inclusiveness} or \enquote{inclusion} specifically when talking about how to include people with disabilities. We acknowledge that this term could and should also consider other systemic issues beyond disabilities and intersections thereof; this is, however, beyond the scope of this paper.

The purpose of this position paper is to raise awareness and spark further discussions on the role of accessibility and inclusiveness for people with disabilities in visualization research. We do not present concrete findings or definite truths, but we want to share our perspective to start conversations and future research efforts.

\section{General Accessibility Research Challenges}
Accessibility research, even beyond visualization research, faces challenges stemming from a more systemic nature.
In this section, we thus want to first discuss the lack of ``disabled voices" within academia. Thereby, we want to raise awareness to potential systemic biases that might leap into our research, making participatory design especially challenging~\cite{Ymous20,campbell2009contours}.

\subsection{How to Provide Equal Opportunities to Disabled Scholars?}
To understand why there is a lack of disabled perspective, let us start with the challenges that disabled academics face.

In the area of human-computer interaction (HCI), Shinohara et al.~\cite{shinohara2020} investigated accessibility issues of doctoral students by conducting interviews. They identified \enquote{\textit{access differentials}}: Their participants described that they often had to put in more time and effort to make progress compared to non-disabled peers, due to inaccessible resources or processes which required workarounds. The services provided did not meet accessibility needs, creating \enquote{\textit{inequitable access}}. The authors concluded that these accessibility issues require systemic change that should be fostered by the HCI community.
Furthermore, Jackson et al.~\cite{jackson2022} reported citational injustice, where accessibility research fails to credit disabled accessibility design experts outside academia, though they provide valuable input to said research.
Ymous et al.~\cite{Ymous20} also criticize systemic discrimination. They point out that even in accessibility research there is still work being published that comes from the ableist perspective and overlooks the disabled perspective. As a disabled researcher researching one's own condition, reading works of others, one often feels \textit{\enquote{dehumanized}}. They also stress the feeling that they, as disabled researchers, often feel that they have to do more work, in particular, to gain access.
Hence they call out for inclusive environment, where accessibility is naturally built-in.

\paragraph{\personal Personal Perspective} I am lucky and privileged to be where I am today. However, it has been hard work to get there. 
I know the feeling of searching for workarounds and inaccessible accessibility solutions that Shinohara et al.~\cite{shinohara2020} describe first-hand. 
Right now, I feel well supported at the institute level and I love my job, even though I do not know how far the academic system allows me to proceed with my lack of mobility. I cannot travel the world multiple times looking for tenure. But one step at a time. In the past, there were always doors that eventually opened.

\paragraph{\solution Potential Solution: Explore the \enquote{new normal}.} The pandemic changed the world as we knew it. The 2020 BELIV workshop~\cite{BELIV2020} 
discussed the impact of COVID-19 on visualization research. Challenges~\cite{balestrucci2020} but also new and adapted evaluation methods were discussed~\cite{losev2020}.
From a general perspective, exploring the ``new normal'' could also have accessibility benefits. Cho et al.~\cite{cho2022} surveyed the benefits and challenges of working from home. They listed benefits such as not needing to commute, more flexibility in planning one's day, and the opportunity to work more focused, at least for some of their participants. These benefits could also be evaluated in an accessibility context. Childhood disability research work by Rosenbaum et al.~\cite{rosenbaum2021let} also reflects on research strategies developed during the pandemic, which they deem as even more accessible and inclusive than processes used before the pandemic.
Whether the same holds true for visualization research remains to be investigated. 

\paragraph{\personal Personal Perspective.}
 For myself, the pandemic had a strange side effect: Regarding my ``disability workarounds", everything was now ``normal", because suddenly the whole world was adapting to exceptional circumstances. The psychological barrier to state my needs was lower and working from home required substantially less scheduling and taking care of accessibility issues. Subtracting those extra 20\% from my 120\% task load was a relaxing experience.

I hope to preserve some of the newly found normality of workarounds and flexibility into the ``new normal'' that we are currently beginning to live.

So, I second Rosenbaum et al.~\cite{rosenbaum2021let} let us not just go back to ``normal'', I hope that we have the courage to actually build a more sustainable and inclusive ``new normal''. 

\paragraph{\call Call for Action: More substantial systemic changes in academia are required.} We are well aware that changing the whole academic system is beyond the VIS community's influence. Nevertheless,  we want to mention this point here for the sake of completeness and to raise awareness. 

\subsection{How to Avoid \enquote{Parachute Research} and \enquote{Disability Dongles}?}\label{sec:frameworks}

A common approach to accessibility is participatory design, a design approach that closely involves end users, in this case disabled users, in the design and iterative evaluation process~\cite{muller1993}. Here, the lack of disabled voices also plays a role.
Lundgard et al.~\cite{Lundgard2019} critically evaluated participatory design efforts in the form of case studies related to tangible visualizations. They warn of participatory accessibility research turning into \enquote{parachute research}, where community expertise is exploited for the sake of publication, but does not really help the community with their research. 
Liz Jackson~\cite{jackson2019} coined the term \textit{\enquote{Disability Dongle: A well intended elegant, yet useless solution to a problem we never knew we had}}, which could be the result of such parachute research. The follow-up article~\cite{jackson2022}, criticizes that assistive technology is often developed for people with disabilities to conform to \enquote{the standard}, rather than developing solutions that would support diverse abilities.

\paragraph{\solution Potential Solution: Go from trying to \enquote{being like} to \enquote{being with}.}Successful participatory design needs an appropriate framework. Here, we are again inspired by the frameworks proposed in HCI.
Constanza-Chock's Design Justice Framework\cite{costanza2020design} states that user-centered design often mirrors power structures of society, and experiences of \textit{\enquote{less privileged}} users are neglected. 
Bennett et al.~\cite{bennett2019} also warn that empathy frameworks might separate and undermine disabled experiences from those of designers. They propose to move from \textit{\enquote{being like}} to \textit{\enquote{being with}}. So, rather than trying to live someone else's experience, the idea is to share experiences together. 

Furthermore, Bennett et al.~\cite{bennett2018} promote the interdependence framework, stating that assistive technology should not be considered as something that bridges gaps between non-disabled and disabled people. Instead, accessibility could be seen as a collective process driven by the relationships between different people, assistive technology, and the environment.

The contribution of Kender and Spiel~\cite{Kender22} outlines pitfalls of participatory design, such as patronizing participants or researcher biases that leap into the design process. 
The response bias in accessibility research is also discussed by Ming et al.~\cite{ming2021}. They identify different dimensions of bias and develop mitigation strategies. 
One should avoid \textit{\enquote{the charity model}}\cite{ming2021} of accessibility research and instead give room for the participant's expertise and experiences. The importance of lived experiences is likewise stressed by McKercher\cite{mckercher2020beyond}. 

We believe, that visualization research could also benefit from such frameworks, either when designing visualizations as assistive technology or when trying to understand accessibility needs during our visualization design processes first hand. Integrating those frameworks might be a first step toward a more ability-aware visualization design, driven by shared experiences.

 \paragraph{\personal Personal Perspective.} 
For me, the first author, my disability is part of my identity and nothing to be fixed. I am perfectly fine with the abilities I have. Please help me fix the inaccessibilities in my environment, though.
As such, my participatory design mantra has always followed the slogan: Nothing about us without us~\cite{Charlton+1998}. With respect to terminology, I think accessible design, where we provide options for people with disabilities, is the point to start with. However, I think the ultimate goal should be inclusive design whenever possible, where we offer accessible options for all from the start. 

\paragraph{\solution Potential Solution: Critically self-reflect on the social impact of your research contribution, to avoid parachute research.} Research can have good intentions in terms of activism, but what is the actual impact of it? De Castro Leal et al.~\cite{leal2021activism} and Liang et al.~\cite{Liang2021} reflect on their experiences and the experiences of fellow researchers on research as activism or allyship. 
They state that not all well-meaning research is actually allyship, but some is in fact harmful. On the other hand, all research should ideally be allyship, especially when working with marginalized people~\cite{Liang2021}. Works like this invite us to self-reflect on our methods and goals asking questions like: What impact does my research have? Does it harm the people with whom I actually want to work? How could I mitigate tensions of allyship?
They encourage a culture of constant critical self-reflection within a research community to counteract potential oppressive structures as a whole~\cite{leal2021activism,Liang2021}.

\paragraph{ \personal Personal Perspective.}
For me personally, research is also a form of social activism. And oftentimes, I read and evaluate research through that lens. Then, the most important evaluation questions are not: Is my visualization significantly better than another? Is my contribution novel enough? But rather: How can my contributions help others? Is this new form of visualization really useful for people? 
Additionally, there is always more than one side to a story. And telling them all is important, as Adichie visualized in her TED talk\cite{adichie2009danger}. She stresses the importance of hearing different perspectives on a topic rather than a \enquote{\textit{single story}}, as this single perspective induces stereotypes, which are an incomplete reflection of the truth. Instead, one has to hear multiple and diverse stories, which then in turn have the ability of empowerment. As Adichie, I also believe that bursting out of one's bubble is an important and educating experience. On the other hand, it is often not that simple. Thus, I often wonder, which single stories bias me and whether my research is inclusive enough.
My personal goal for my work is to help me and others see our research and perhaps even society through different perspectives. 

\section{Accessibility Challenges in Visualization}

Besides the above systemic challenges, thoroughly considering accessibility concerns during visualization evaluation and design brings challenges on its own. Those challenges are mainly inspired by our own research experiences so far and are by no means complete. They mostly focus on accessibility during the evaluation pipeline, but also briefly touch on tensions in visualization design. 

\subsection{How Accessible Are Study Setups?}
One key question when planning a study is: Is my setup suitable?
In accessibility research, another key question is: Is my setup accessible enough for my target audience? 
There are many potential roadblocks that could prevent disabled people from participating in visualization experiments, such as inaccessible hardware or the lack of assistive technology~\cite{petrieremote2006}.

\paragraph{\solution Potential solution: Remote setups could be more easy to participate in, at least for some people with disabilities. } Remote accessibility studies are not a new phenomenon, but have been investigated already by Petrie et al. in 2006~\cite{petrieremote2006}. They see remote studies as a possibility to reach more participants and reduce the entry barrier to a user study. Since the start of the pandemic, the number of general remote study setups have increased, such as remote virtual reality (VR) studies~\cite{Siltanen21,radiah21}. Likewise, in the domain of accessibility research, evaluation had to become remote or virtual~\cite{liedke2021,lee2022}.

Although mostly due to the pandemic, all of our accessibility-related studies became remote studies. Although this presented challenges, such as lack of control over the experiment environment~\cite{Kittur2008}, it also provided the opportunity to reach people from all over the world and from the comfort of their own homes. To reinforce the point of Petrie et al.~\cite{petrieremote2006}, this was especially useful for reaching people with disabilities in our case people with limited mobility or limited color vision. It is difficult to build a local network of disabled participants that meets all the requirements of your study~\cite{petrieremote2006}. If one can expand it to the whole world, it becomes at least somewhat easier. 

Furthermore, in the event of a pandemic, remote studies do not put your participants' health at risk, which is even more critical for people in high-risk groups. 

However, as all accessibility matters, this is not meant as a \enquote{one size fits all} solution, since remote setups might be more accessible to some but at the same time more cumbersome to others. Furthermore, it could also depend on the complexity of the study setup, whether or not it is suitable to be done remotely.

\paragraph{\personal Personal Perspective.}
Speaking from personal experience, it is easier to conduct an online interview from home than to supervise a study in the lab. This could also be true for some of my participants, who may not have been able to join my studies not only due to geographic restrictions but also because it might be too inacessible. 

\paragraph {\solution Potential Solution: Crowd workers might have intrinsic motivation to do accessibility tasks.} Crowdsourcing platforms provide another means for remote evaluation. Since crowd working platforms have been discovered as a means of evaluation in HCI and visualization research~\cite{Kittur2008,Heer2010}, they have also been used for accessibility tasks such as generating image captions~\cite{Simons2020} or assessing the accessibility of sidewalks and streets~\cite{Saugstad2019,Hara2013}.
Simons et al.~\cite{Simons2020} reported that crowd workers were sometimes insecure regarding the helpfulness and correctness of their answers, but also motivated to do accessibility tasks.

In our own previous work~\cite{angerbauer2022}, we also used crowd-sourcing to evaluate accessibility data visualizations with respect to color vision deficiencies. 
Echoing related work, we discovered a certain insecurity among crowdworkers, whether their answers were correct and helpful, and our study required higher communication efforts than other crowd-sourcing studies we had conducted before. At the same time, we noticed high intrinsic motivation for accessibility tasks among crowd workers. 

\paragraph{\call Call for Action: We need to develop ways to ensure accessibility of study procedures.}
In the past, we surveyed, evaluated, and compared research methods regarding the validity of the results that different evaluation methods produce~\cite{Isenberg:2017:VMC,voit2019online,weiss2020revisited,isenberg2013systematic} 
To the best of our knowledge, there is little work on a systematic assessment of the accessibility of existing evaluation methods. If we want to strive to make visualization design and evaluation truly inclusive for people with disabilities, this would be a viable route for future work. 
Some questions as a starting point: What are accessibility hurdles of our study procedures and how can we mitigate them? What study methods are best suited for limited mobility, low vision, or any other disability?

\subsection{How to Recruit and Include Your Target Audience to Evaluate Accessibility?}

After deciding on a study setup, the next step is to recruit members of the target audience, with which to evaluate your visualization. However, dealing with a small number of participants appears to be a recurrent issue in accessibility research~\cite{sears2011,mack2021}. As mentioned above, it is difficult to build up a local participant pool~\cite{petrieremote2006} and participants with disabilities are often underrepresented.

\paragraph{\solution Potential Solution: A small number of participants can provide meaningful results.} 

 Mack et al.~\cite{mack2021} argue that even small $n$ in accessibility research should be considered meaningful and that one should be careful to fill the participant pool with non-disabled participants to not reinforce ableist biases. Working with small participant samples $n$ is also common in other visualization studies, specifically when domain experts are needed~\cite{sedlmair2012design,isenberg2013systematic,shneiderman2006strategies,plaisant2004challenge,carpendale2008evaluating}. As such, there is some expertise in the visualization community in working and evaluating with scarce user groups. So far, this expertise has rarely been used for the target audience of disabled people, but might offer good starting points to be leveraged for accessibility studies.

\paragraph{\solution Potential Solution: Be creative with respect to recruitment channels.} In our research to date, we also struggled with the underrepresentation of disabled users. 

Studies, for which we actively thought to recruit disabled participants, participant recruitment was much more time-consuming than expected, and we had high dropout rates when recruiting through social media. Still, we believe that going different recruitment routes in addition to mailing lists could be fruitful. Posting on general channels of social media might induce not-so-serious participations; however, when locating disability-related channels or personal profiles of disability advocates, one might be more likely to reach desired participants. Even if disability advocates may not be able to join your study themselves, they could boost the invitation through their channels, hopefully increasing the turnout. 
Another channel to connect with participants is by getting in touch with local communities. In his 2020 ASSETS keynote, Lazar~\cite{lazar2020accessibility} recommended to reach out to local disability services to connect with participants and their needs. 
Thus, printing paper flyers and handing them out to disability services at the university or your local community might be beneficial. Additionally, on crowdsourcing platforms, one might even be able to pre-select participants with certain disabilities.

\subsection{How to Explain and Define Accessibility? }\label{sec:explain}
Both for explaining the accessibility evaluation task to the previously recruited participants as well as to communicate accessibility needs to visualization designers, we need clear guidelines.
In the field of web accessibility, there exist specific guidelines, such as the Web Accessibility Guidelines (WCAG)
\cite{WCAG}, and there are metrics for visualization linting ~\cite{mcNutt2020mirages,hopkins2020visualLint}. However, there are no clear definitions for the accessibility of visualization. Designers are often confused by the individuality~\cite{Joyner2022} and so are participants according to our own experiences.
Explaining accessibility might require greater communication efforts~\cite{angerbauer2022}.
We need to find a way to provide guidelines and account for individual exceptions when appropriate to fully grasp the concept of accessibility, which makes the definitions of simple metrics challenging.

Another challenge related to the definition of accessibility is the lack of datasets about it and the fact that data often lacks the perspective of people with disabilities in general. We consider the following data as accessibility datasets: all data generated by people with disabilities either on accessibility itself, on disability-specific topics, or data to provide a perspective from disabled people on general design ideas. These data sets, such as a collection of different sign language gestures~\cite{dreuw2006modeling} could inform machine learning or design processes.
\paragraph{\personal Personal Perspective.} 
I have Cerebral Palsy. A friend of mine does too, but we have different abilities. Similarly, I use a walker like some elderly, but my sense of balance is worse. Accessibility needs are not simply defined. Diagnosis and aids can be an indicator, but also misleading. To ultimately gain knowledge about user needs, we have to start a conversation with them. 
\paragraph{\solution Potential Solution: Utilize simulators (with care).}

Simulators~\cite{Goodman-Deane2007,krosl2020cataract,Schulz2019,Aytac2017,Flatla2012} can help visualization designers, but should be used with care, as they may encourage the criticized \textit{\enquote{being like}} attitude~\cite{bennett2018,tigwell2021}, and fail to account for the individuality of different disabilities.
\paragraph{\call Call for Action: We need to investigate the sweet spots and danger zones of simulator use further.} 
Simulators were helpful in our previous project~\cite{angerbauer2022}, however, they do have the limits described above. A thorough and critical investigation of the situations in which simulators are helpful and those in which they are not might help to judge risks and potentials.
\paragraph{\personal Personal Perspective.} From my point of view, I would  not use simulators and smart tools to fully replace people in your accessibility evaluation. You might simulate some aspects of my disability, but you cannot simulate me and my experiences.
Simulators provide you with a starting point for conversation and further evaluation, no ground truth. With that in mind, I think simulators or even mixed reality could provide interesting technical means to build bridges between perspectives relevant in visualization design and evaluation. 

\paragraph{\solution Potential Solution: Use heuristics as a starting point o explain accessibility.}
Another way to evaluate accessibility is through heuristics, for instance color vision deficiency heuristics~\cite{martinez2021methodology}.
Recent work by Elavsky et al.~\cite{Elavsky2022} approaches broader accessibility heuristics for visualizations with their Chartability heuristics, factoring in vision, but cognition and motor abilities needed to understand and interact with visualization.

\paragraph{\call Call for Action: We need more accessibility datasets related to visualization research.}

Within HCI research, accessibility datasets have recently gained increasing attention: Mack et al.\cite{mack2021} called out for more accessibility datasets, to avoid biases in artificial intelligence and Theodorou~\cite{Theodorou2021} provided guidance on disability-first data collection. IncluSet~\cite{kacorri2020incluset} is a starting point for more accessibility datasets: It provides information on 139 related datasets, some directly downloadable and some available on request. However, its 31 image datasets directly available on the website\footnote{\url{https://incluset.com/datasets}} are not directly related to visualization research. Such datasets, we believe, could inform visualization design decisions in favor of accessibility and inclusion. Among other things, we need datasets where people with disabilities judge the accessibility of a visualization in the light of their disability or where interactions with visualizations by people with disabilities are recorded to uncover potential stumbling blocks in the interaction design. 

Therefore, the VIS community should foster and encourage the creation and publication of accessibility datasets while ensuring ethical standards, which are important within participatory design~\cite{Kender22}. Here , appropriate tooling and infrastructure for collecting accessibility annotations could be developed. 

\paragraph{\call Call for Action: We need tools to facilitate \enquote{sharing of experiences}.}
Related to the above call for action to support the collection of accessibility datasets, we need tools and frameworks that enable people with disabilities to tell their stories authentically, while designers listen. This goes beyond using simulators, although they could be a starting point. An example from HCI is a realistic wheelchair simulation in virtual reality, where there is actually a participant with a disability who uses the simulator in a wheelchair, while consulting architects on the development of a smart city~\cite{goetzelmann20}.
For visualization research, we could build shared-experience evaluation frameworks, where designers get pointers by leveraging knowledge from previously created accessibility datasets in addition to using simulators.

\subsection{How to Evaluate and Report Accessibility Results?}
Even if we have accessibility data available, we finally have to decide on the manner of evaluation and reporting.
As mentioned above, accessibility is highly individual and it may become difficult to distinguish subjectiveness from noise and find ways to honor individuality within the collected data while evaluating and reporting accessibility results ~\cite{angerbauer2022}.

\paragraph{\solution Potential Solution: Outliers could be meaningful.}

In theory, we have standardized methods to deal with outliers in our user study data~\cite{aguinis2013best}. However, excluding too many data points could also lead to selection bias~\cite{peer2014reputation}. Furthermore, what if the outlier could tell us something meaningful? Sometimes usability issues that one person has might be interesting to the designer~\cite{asborn2012}.
We believe that this fact might especially hold true for accessibility research. As accessibility is highly individual, ideally every opinion should count.
In our previous work~\cite{angerbauer2022}, we excluded as few outliers as possible and provided multiple perspectives on the data through a minority and majority analysis.

\paragraph{\call Call for Action: We need to visualize diversity of our accessibility results and provide different interpretations for our data.}
More research is needed on how to best visualize diversity of our results, as also noted by Schwabish and Feng~\cite{visCommPaper2020Schwabish} in the context of creating racial equity-aware visualizations. We also need guidelines on how to visualize data for people with diverse abilities. Multiple data narratives could not only help accessibility research, but also make research more transparent in general~\cite{dragicevic2019increasing}. 

\paragraph{\call Call for Action: Dare to report your challenges and potential faults.}

The paper we wanted to write about accessibility of paper figures~\cite{angerbauer2022} initially had no ``challenges" in the title. When beginning to design the data study, we expected to have a clear overview of the accessibility of paper figures and some solid design guidelines. What we actually got was quite fuzzy data. However, the challenges we faced were interesting, so we decided to report them to help other researchers with similar endeavors. 

\paragraph{\personal Personal Perspective.}
It often feels like we are trained to write our research as success stories and to somehow sell our shortcomings as benefits.
I believe that all the challenges I faced made me stronger and the person I am today. On the other hand, I learned that we also need to communicate when we struggle and need help to truly succeed. 
Part of research is the failure to meet expectations. But if we do not report our failures, how can others know about them and learn or avoid making the same mistakes again? For me, the most exciting thing is to debug why something did not work as expected.

\subsection{How to Deal With Tensions Between Design Choices and Accessibility Needs?}\label{sec:design}
Besides taking care of accessibility in the evaluation process, we also need to aim for accessibility during the design process. However, this is not easy in some cases as there could be trade-offs between design dimensions and accessibility needs. Here is one example: 
According to Joyner et al.~\cite{Joyner2022} but also our own experiences, simple visualizations often are the most accessible ones. The more complex they are, the lower their accessibility. As such, there is a tension between the visualization designer's aim of creating complex, rich visualizations and the requirement of making them accessible. Some designers would not sacrifice richness of information for accessibility~\cite{Joyner2022}.

\paragraph{\solution Potential Solution: Consider accessibility from the start. } We agree with Joyner et al.~\cite{Joyner2022} that visualization accessibility should not only be considered in retrospect, but as one of the design goals from the beginning.
One way to do this could be with the shared-experienced-driven design processes discussed in Section~\ref{sec:frameworks}.
\paragraph{\call Call for Action: We need to investigate accessibility trade-offs.}
To mitigate tension, we need to know what kind of tensions there are. These also might be tensions to other design dimensions such as information richness, interactivity, aesthetics, and potentially many more.

\paragraph{\personal Personal Perspective.}
Accessibility is not always attractive, making it even harder to sell. There are people who complained about expenses or workarounds they had to take on my account. I am thankful, but I am not apologizing anymore.
Yes, textures or accessible color schemes might seem \enquote{ugly} to some, but I will continue using them. Accessibility and functionality have always won over aesthetics in my life. 
But I acknowledge the temptation to go the easy route, especially when time is limited. Hence the fervent call for tool support.

\section{Discussion and Limitations}
In this position paper, we only discussed being disabled as one factor of inclusion. Of course, there are also other factors, such as ethnicity, age, and gender, that we need to consider for a truly inclusive design beyond disability. In addition, our key points are mostly borrowed from the general HCI literature. We do not have many explicit examples from visualization research, simply because the literature on accessibility within the VIS community is still scarce.
Furthermore, our solutions are only a starting point toward the goal of inclusive evaluation and design processes. 
To finally arrive at inclusive visualization evaluation and design, we still have a long road ahead of us.

Tackling the raised issues in an isolated manner, as for instance by broadening one's participant pool but still relying on narrow, simplified definitions of accessibility, might still result in non-inclusive visualizations. If we want to approach inclusion, we need to do it holistically. How could we achieve this? We do not have an overall solution, only speculations. 

For one, we have to acknowledge that our research views might inherently be biased by existing societal structures, and that the needs of (intersectional) minorities might be easily overlooked. 
By acknowledging biases within our development and research goals, we can find solutions on how to mitigate them. As mentioned above, that might be uncomfortable at times and even more costly.
But we should ask ourselves questions like: Who benefits from our visualization research and who does not? Is this disadvantage intentional (it might be the case for expert tools) or is the disadvantage actually stemming from inherent biases that caused us to neglect certain groups? To achieve increased replicability of our papers, there exists the replicability stamp\footnote{http://www.replicabilitystamp.org/}. Do we want to create a diversity/inclusivity stamp as well, which could help to evaluate our research from a more systemic perspective?
Furthermore, we should actively engage with marginalized communities and foster community-driven projects \cite{costanza2020design}.

There might be more than one way to reach the goal of accessibility and inclusion in visualization research, which also is an essential message of diversity.
While striving for the overall systemic solution, which goes beyond visualization research, at the lower level, we might have to say goodbye to the \textit{\enquote{one size has to fit all}} solutions and be more creative to accommodate diverse needs. Inclusive design in some cases might mean offering the same beneficial functionalities for all, in other cases it might mean offering basic common functionalities plus additional individual functions to be enabled as needed. A non-visualization example for the first case: Ways without stairs might benefit not only people with disabilities, but also the general public. For the second example: We might have a basic simple visualization as our main prototype, which could have additional expert/novice options and additional accessibility options like texture or read-aloud functions. 
When to choose which options should be carefully considered. Certainly, there are limits of accessible design. But we assume that there are also benefits to it, which we have not even started to take advantage of, since accessibility has been largely neglected in the past.

Another key point to tackling accessibility and inclusion is to start to understand it in the first place. As stated in Sections~\ref{sec:explain} and~\ref{sec:design}, accessibility is individual and difficult to grasp, and trade-offs in design are under-researched. As such, the challenges in those sections are strongly connected. Once we have reached a definition of what accessibility is in particular use cases, we can investigate trade-offs and adapt our choices accordingly. To achieve this understanding, the participatory design frameworks mentioned in Section~\ref{sec:frameworks} can allow shared experiences and perspective exchange. 

In our speculations, one advantage of visualization research is that it in fact can help itself in fostering inclusion. Visualization can aid in the communication of facts and experiences. We can create tools that visualize our different experiences to communicate our access needs and raise awareness. 
To get rid of biases and to understand accessibility, communication, and perspective exchange are key in our opinion. With visualization research, we could eventually provide a toolbox for that exchange.

\section{Conclusion}
This position paper discussed the topic of accessibility and inclusion in visualization research, illuminating it from different perspectives. We looked at underlying systemic aspects and thought about the difficulties and chances we face when striving toward a more inclusive visualization design and evaluation process.
However, we want to stress that even though we reported accessibility from different angles, it is still seen through our lens, and by far not complete. 
Disabled researchers are also not immune to being biased~\cite{hofmann2020}.

Besides striving for more accessible and inclusive visualizations in particular, we believe that we should re-inspect our well-established methods again in light of inclusivity to avoid propagating biases accidentally. 
As Hofmann et al.~\cite{hofmann2020}, we believe that there is a strength in creating connections between different perspectives, as this has the potential to create more inclusive (research) world. Creating connection could mean to on the one hand, to foster the connection between disabled accessibility researchers or accessibility researchers in the VIS community, and, on the other hand, to connect different experiences of people with diverse needs. 
With this approach, we could potentially consider accessibility from the beginning with universal design, rather than simply providing patches for \enquote{able-experience}. In the long run, ideally, we might even leverage each user experience as equally important.
To achieve this goal, looking at our research from multiple perspectives might help. We encourage everyone, including ourselves, to change their perspectives once in a while. And we see visualization research in the position of providing tool support for that matter.

\acknowledgments{
This work is funded by the Deutsche Forschungsgemeinschaft (DFG, German Research Foundation) – Project-ID 251654672 – TRR 161.}

\bibliographystyle{abbrv-doi}

\bibliography{template}
\end{document}